\documentclass{article}

\usepackage{PRIMEarxiv}

\usepackage[utf8]{inputenc} % allow utf-8 input
\usepackage[T1]{fontenc}    % use 8-bit T1 fonts
\usepackage{hyperref}       % hyperlinks
\usepackage{url}            % simple URL typesetting
\usepackage{booktabs}       % professional-quality tables
\usepackage{amsfonts}       % blackboard math symbols
\usepackage{nicefrac}       % compact symbols for 1/2, etc.
\usepackage{microtype}      % microtypography
\usepackage{lipsum}
\usepackage{fancyhdr}       % header
\usepackage{graphicx}       % graphics
\graphicspath{{media/}}     % organize your images and other figures under media/ folder
\usepackage{tabularray}
\usepackage{pdflscape}
\usepackage[longtable]{multirow}
\usepackage{longtable}
\title{Privacy's Peril: Unmasking the Unregulated Underground Market of Data Brokers, and the Suggested Framework}
\author{
  Rabia Bajwa \\
  MCTI Program  \\
  University of Guelph \\
  Guelph\\
  \texttt{Rbajwa01@uoguelph.ca} \\
   \And
  Farah Tasnur Meem \\
  MCTI Program  \\
  University of Guelph \\
  Guelph\\
  \texttt{ftasnurm@uoguelph.ca} \\
}

\begin{document}
\maketitle

\begin{abstract}
The internet has made it a common place for businesses to gather and store as much customer data as they can, and computer storage capacity has increased exponentially in tandem with this trend. Businesses use this data to enhance customer satisfaction, grow revenue, boost sales, and increase profitability. Nonetheless, the growing field of Data Brokers is still fraught with legal difficulties. We will go over what a Data Broker is, how it gets information, the data industry, and some general issues it deals with in Part I. The various approaches to regulating Data Brokers will be examined in Part II; all strategies are suggested with consideration for the EU GDPR. In Part III, we will provide our own analysis and conclusions in addition to providing more responses to the Data Broker industry's worries.
\end{abstract}

% keywords can be removed

\section{Introduction}
Data Brokers are not a novel idea; they have been around for a while. Data Brokers gathered most of their information offline in the 1960s. Personally identifiable information ("PII") is a category of information that Data Brokers gather and is described as "information that can be used to distinguish or trace an individual's.

\subsection{Who Data Brokers Are?}
An organization that gathers, analyzes, and licenses user's personal information to be used by other businesses for purposes such as marketing is called a Data Broker, sometimes referred to as an information product company. "\cite{r8}There are many different types of Data Brokers; some work directly with clients, while others do not.

\subsection{How Do They Collect The Personal Data?}
Data Brokers gather information from three primary sources: publicly accessible sources, government records, and commercial sources. \cite{r9} Government sources include federal, state, and local records. These records may contain a person's name, address, age, political affiliations, marital status, family structure, property deeds, criminal histories, income and tax records, and licenses. \cite{r10} Information from blogs, social media, and other websites with no privacy settings is included in publicly accessible sources.
\cite{r11}These two less-regulated categories allow Data Brokers to collect information from individuals without their consent. The third commercial source contains information obtained through personal activities such as account registration or online purchases. People typically explicitly consent to the terms and conditions that allow the collection and use of their data when they sign up for services or accounts. 
\begin{figure}[htbp]
\centering
  \includegraphics[width=0.45\columnwidth]{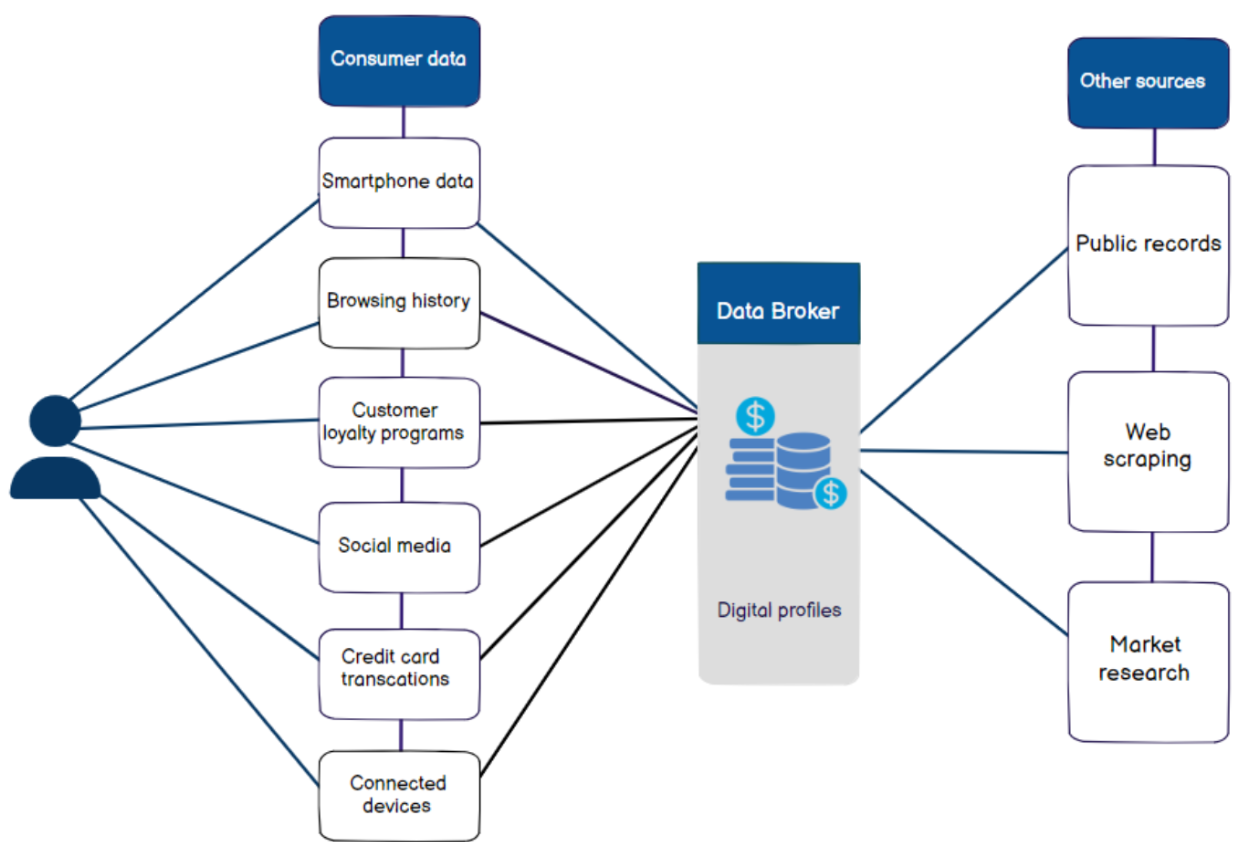}
  \caption{Data Brokers gathering data}~\label{fig:figure1}
\end{figure}

The multitude of sources from which Data Brokers gather their data is illustrated in the following figure \ref{fig:figure1}. 

\subsection{What Data Brokers Do With Personal Data?}
Data Brokers combine offline and online information, such as voter lists and home purchase records, to build comprehensive profiles of specific people. \cite{r17}By combining online and offline data, including voter lists and public records like home purchases, Data Brokers are able to create comprehensive profiles of specific individuals.\cite{r18} This method, known as the "aggregation effect," combines raw data to infer behaviors and personality traits. 
\begin{figure}[htbp]
\centering
  \includegraphics[width=0.45\columnwidth]{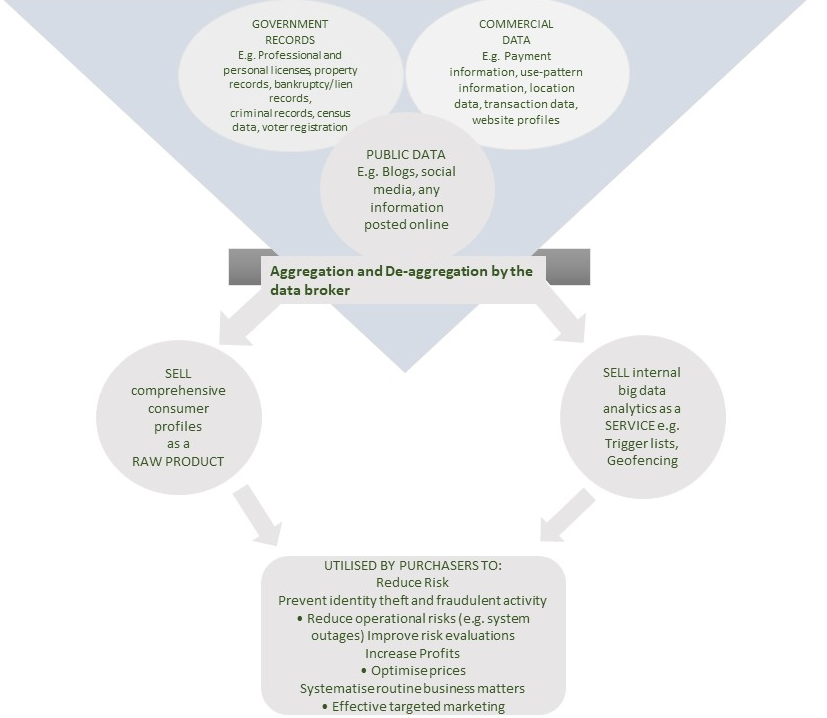}
  \caption{Aggregation and De-aggregation by the Data Brokers}~\label{fig:figure2}
\end{figure}
The aggregation and deaggregation by Data Brokers is illustrated in figure \ref{fig:figure2} This information is primarily used for marketing and advertising. 

\begin{figure}[htbp]
\centering
  \includegraphics[width=0.45\columnwidth]{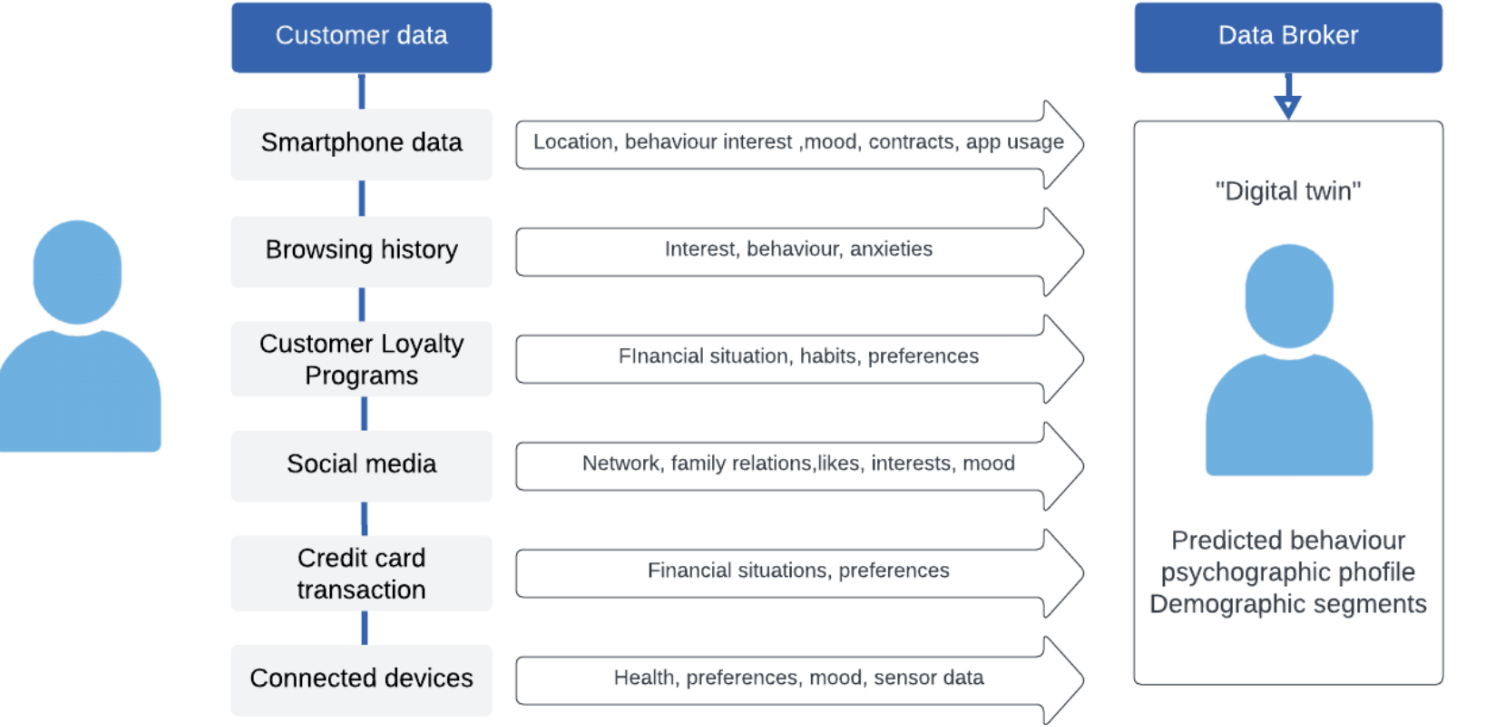}
  \caption{The making of your ‘Digital Twin/Dossier’}~\label{fig:figure3}
\end{figure}

More troubling is its use in generating a "propensity score," a non-financial credit score based on behavioral and other factors that employers and businesses use covertly to assess dependability.[9] This extensive data collection and analysis enable brokers to construct a "digital twin" or dossier of individuals is illustrated in Figure \ref{fig:figure3}

\section{Related Literature Review}
There have been numerous studies conducted in the field of GDPR and Data Brokers, and some of the ones we considered are listed below:
The study "The Dark Industry of Data Brokers: Need for Regulation?" identifies legal gaps that allow Data Broker operations and advocates for regulatory reforms, particularly in countries such as India, where data protection laws are in their infancy \cite{r1}.

The article "Guidelines for GDPR Compliance in Big Data Systems" addresses the GDPR's new security and data privacy regulations. It emphasizes how important it is for businesses to abide by these rules in order to safeguard rights and enhance business models. \cite{r2,r3}.

The article "Data Protection, Scientific Research, and the Role of Information" evaluates the GDPR's information obligations, particularly when it comes to scientific research settings \cite{r4}.

The challenges of integrating blockchain technology with GDPR regulations are the main topic of "GDPR Compliant Blockchains: A Systematic Literature Review" \cite{r5}.

Another study on "Guidelines for GDPR Compliance in Big Data Systems" indicates that GDPR discussions in academic and industry papers have covered a variety of perspectives, but that these works are still in their early stages.
 \cite{r6}.

“Guidelines for GDPR compliance in Big Data systems” examines the effectiveness of transparency in addressing privacy concerns in the Data Broker industry, arguing for a shift toward a commodification-based approach to shaping public policy \cite{r15}.

“The ChoicePoint Dilemma: How Data Brokers Should Handle the Privacy of Personal Information” summarizes the 2005 ChoicePoint data breach, highlighting the exposure of personal information as a result of unauthorized access. The paper appears to focus on the consequences of the breach from various perspectives, such as consumers, executives, policy, and IT systems, while also making future recommendations \cite{r16}.

\section{Methodology}
Using a mixed-methods approach, case studies on Data Brokers, data handling analysis, and a fictitious framework for compliance with data privacy and protection (mainly in line with GDPR requirements) are all included. We examined a range of data breach incidents and, in certain cases, the potential for data breaches or privacy control violations had the applicable privacy frameworks been in place. We also located records of Data Brokers that demonstrate data breaches that have occurred over the previous 20 years, which amply demonstrates the serious shortcomings in data protection and privacy that Data Brokers have control over. The study concludes with actionable suggestions for Data Brokers to enhance their data security and privacy procedures, based on the suggested conceptual framework. Further details of contributing elements towards research are highlighted below:

\begin{itemize}
    \item \textbf{Selection of Subjects:} 
    
\textbf{Large Data Brokers and people-finding websites:}

We identified 09 major Data Brokers after a meticulous compilation process that was aided by research papers and databases such as: Privacy Rights Clearinghouse | Data Brokers. .
\textbf{Material and documentation review:}

Public Information Analysis: We carefully examined publicly accessible records, including marketing and promotional materials on their websites, to learn more about the Data Brokers' methods for promoting personal information about the people for whom the data was gathered. 

\end{itemize}

\begin{itemize}
    \item \textbf{Methodological Approach:}
    
Case Studies Analysis: To gain a thorough understanding of each selected Data Broker's operations, detailed case studies were considered.
Compliance Assessment: To assess the GDPR adherence level of these Data Brokers, a hypothetical Data Privacy and Protection program mapped with GDPR requirements has been developed and proposed. We also, considered studying the requirements given under NIST Privacy Framework, CSF Privacy Controls, OECD's Data Protection Rules, and ISO/IEC 27701:2019.
Tools Used: We have used Excel, Visio, and Xmind map for the development of graphs, figures, and mapping flows respectively. \end{itemize}

\section{Problem statement}
The Data Brokerage practice, which includes larger purchases, licensing, and sharing of data, as well as the Data Brokerage industry, which includes Data Broker firms specifically, are essentially unregulated due to the lack of virtual controls for data privacy and protection of the information that these businesses collect and process.

This research investigates how Data Brokers collect user data and, because they are not subject to any privacy laws or standards, what obligations they have by disclosing information about Data Broker data breaches and the consequences that result.The study's objectives are to:

\begin{enumerate}

\item Despite the financial benefits, the Data Brokerage industry faces serious problems with consumer security, privacy, and potential abuse. The frequent gathering and distribution of erroneous or false information is a serious issue. Consumers may suffer if decisions are made based on inaccurate data, even though businesses may find value in all information, regardless of its accuracy. This can result in a number of unfavorable outcomes, including lower credit scores, higher product rates, or benefit denials. As a result, there is an increased need for stricter controls and accurate data collection and distribution by Data Brokers \cite{r12}.

\item The issue is that identity theft incidents may arise as a result of risk mitigation strategies in the Data Brokerage industry failing. Under these circumstances, fraudsters can use fake information to pretend to be authentic clients. When discrepancies between the fraudulent and authentic data are discovered later, these innocent clients are mistakenly classified as fraudulent. In order to safeguard customers from the consequences of identity theft and incorrect classification, this scenario emphasizes how urgent concerns about data integrity and ethical use in the Data Brokerage sector must be addressed\cite{r21}.

\item The problem arises from the practice of categorizing people based on potentially erroneous or misinterpreted data, which is becoming more and more common in the Data Broker sector. People are categorized through this process according to a variety of assumed characteristics, including gender, income, spending habits, race, and ethnicity. Businesses and organizations use this data-driven classification for things like determining insurance rates and determining a student's ability to repay student loans, but it frequently makes incorrect assumptions. These conclusions are based on a person's online activity and purchases, which may not fully represent their real situation or potential. As a result, important decisions affecting sectors like healthcare and education are made on the basis of possibly inaccurate information\cite{r20}.

\item The main issue in the Data Broker industry is that consumers do not have enough control or transparency over their personal data. The lack of strict laws or regulations governing these brokers aggravates the problem by making it difficult to verify the accuracy of the data gathered and processed. As a result, consumers are frequently forced to make critical—and sometimes life-altering—decisions based on inaccurate or misleading data that they are unaware is being used in such critical situations. A health insurance company, for example, may use an individual's online search and shopping patterns to calculate their insurance premiums—often without the policyholder's knowledge or consent and based on potentially incorrect information\cite{r19}.

\item There are insufficient safeguards in place in the current Data Brokerage system to prevent dishonest people from manipulating politics, the economy, or society using purchased data. Facebook works with Data Brokers even though it is not one; in order to improve its targeted advertising, Facebook sells ads to accounts connected to Russia, which has been known to spread misinformation and incite conflict in the United States. This case emphasizes how dangerous it is for malicious actors to misuse the system in order to harm innocent people, and how Data Brokers have no moral or legal duty to stop this kind of abuse\cite{r13}.

\end{enumerate}

\section{Objectives}

\textbf{To Analysis the Impact of Data Handling Practice in Data Brokerage Industry:} 

 The purpose of the study is to look into and analyze the effects of Data Brokerage practices by looking at data breach incident history. This may entail examining the effects of inaccurate data collection and dissemination on customers, investigating cases of identity theft connected to Data Brokerage missteps, and assessing the practical effects of data classification techniques employed by brokers on important life choices. 

\textbf{To Discuss Most Prominent  Data Privacy and Protection Laws: }

We wanted to study and examine what are the data privacy and governance framework available across the globe that enforces controls toward users' PII. Some of them that we considered included GDPR, PIPEDA, FCRA, GLBA, HIPAA, COPPA. All of these laws together set important guidelines for responsible data governance, demonstrating a global commitment to protecting people's privacy in the digital age.

\textbf{Considering the Data Privacy Impact propose a framework:} 

The study aims to look into and analyze the effects of Data Brokerage practices by looking at particular instances and situations in the sector. This entails examining the effects of inaccurate data collection and dissemination on customers, investigating cases of identity theft connected to Data Brokerage missteps, and assessing the practical effects of data classification techniques employed by brokers on important life choices. 

%\chapter{ \textbf{PART II}}
\section{Most Prominent Laws}
We have illustrated Data privacy law and their 
The most prominent laws are illustrated in table \ref{tab:laws}
\begin{longtblr}[
  caption = {Most Prominent Laws},
  label = {tab:laws},
]{
  width = \linewidth,
  colspec = {Q[56]Q[142]Q[96]Q[644]},
  cells = {c},
  cell{2}{1} = {r=5}{},
  vlines,
  hline{1-2,7-9} = {-}{},
  hline{3-6} = {2-4}{},
}
\textbf{Count- ry Region} & \textbf{Data Privacy Law} & \textbf{Applicable to} & \textbf{Key Features and Details}\\
\textbf{United States} & Sectoral Approach & Data Brokers, various industries & This approach, which combines federal and state regulations, does not have a single national standard created specifically for Data Brokers. It gives industries the freedom to innovate, but it provides only a limited level of comprehensive protection for sensitive personal data. Regulatory oversight is primarily the responsibility of the Federal Trade Commission (FTC).\\
 & Federal Credit Reporting Act (FCRA) & Consumer reporting agencies, Data Brokers & The FCRA regulates the privacy and accuracy of data in consumer reports. It limits the data that can be included in these reports, excluding things like personal information and out-of-date negative information. The legislation also makes a distinction between Data Brokers and consumer reporting agencies; the latter are usually not involved in assessing a consumer's eligibility.\\
 & Gramm-Leach-Bliley Act (GLBA) & Financial institutions & The regulation of the gathering and sharing of personal financial data is the main goal of GLBA. Financial institutions must provide consumers with the opportunity to opt out of having their personal information shared with third parties, as well as clear privacy notices. The act restricts the disclosure of nonpublic personal data to unaffiliated third parties.\\
 & Health Insurance Portability and Accountability Act (HIPAA) & Healthcare providers, health insurers & Sensitive patient health information is shielded from disclosure without the patient's knowledge or consent thanks to HIPAA. It primarily affects insurers and healthcare providers by establishing standards for the security and privacy of health information. Health apps and tech companies are examples of non-covered entities that face difficulties when collecting health data.\\
 & Children's Online Privacy Protection Act (COPPA) & Websites collecting data on children under 13 & COPPA governs how websites and online services can obtain personal information about children under the age of 13. It offers strict enforcement of compliance and necessitates getting parental consent for data collection. Its applicability is restricted to websites that intentionally gather information from minors, so some websites are able to get around its rules.\\
\textbf{Europ- ean Union} & General Data Privacy Regulation (GDPR) & All businesses collecting data on EU citizens & GDPR is a comprehensive data protection law that is applicable to all companies that handle the personal information of individuals living in the EU. It places a strong emphasis on accountability, transparency, and people's rights regarding their personal data. Important clauses cover data portability, the right to be forgotten, the right to access personal data, and stringent consent requirements. severe penalties for breaking the rules.\\
\textbf{Cana- da} & Personal Information Protection and Electronic Documents Act (PIPEDA) & Private sector organizations & Canada's private sector organizations are subject to the Personal Information Protection Act (PIPEDA), which regulates the gathering, use, and disclosure of personal data for commercial purposes. It places special emphasis on the right of individuals to access and update their personal data, as well as consent for data collection and processing with a legitimate purpose. The security of personal data is covered by PIPEDA as well.
\end{longtblr}

The United States takes a sector-specific approach to regulating data privacy, enacting different laws for different types of industries. Data Brokers navigate a patchwork of federal and state regulations under the direction of the Federal Trade Commission (FTC), operating in the absence of a single national standard. This strategy only partially protects sensitive personal data, even though it allows for industry innovation. The privacy and accuracy of consumer reports are addressed by laws like the Federal Credit Reporting Act (FCRA), which makes a distinction between consumer reporting agencies and Data Brokers. With a focus on financial institutions, the Gramm-Leach-Bliley Act (GLBA) mandates unambiguous privacy notices and gives customers the choice to "opt out" of data sharing. Patient health information is protected in the healthcare industry by the Health Insurance Portability and Accountability Act (HIPAA). Furthermore, the Children's Online Privacy Protection Act (COPPA) governs websites that collect data on children under the age of 13, emphasizing strict compliance and parental consent for data collection in limited circumstances. 

The European Union General Data Protection Regulation (GDPR) is a comprehensive data protection law that applies to all businesses that collect data on EU citizens. GDPR introduces key provisions such as data portability, the right to be forgotten, and the right to access personal data, with the goal of improving accountability, transparency, and individuals' rights concerning their personal data. The regulation imposes stringent consent requirements and carries severe penalties for violations, emphasizing the importance of protecting personal information. The GDPR's comprehensive framework aims to ensure that companies handling personal data of EU residents adhere to strict standards, promoting a higher level of data protection across various industries. 

In Canada, private sector organizations are governed by the Personal Information Protection and Electronic Documents Act (PIPEDA), which governs the collection, use, and disclosure of personal data for commercial purposes. Individual rights are prioritized in this legislation, particularly the right to access and update personal information. PIPEDA requires consent for the collection and processing of personal data, as well as a legitimate reason for such actions. Furthermore, the Act addresses personal data security, emphasizing safeguards for sensitive information. Overall, PIPEDA creates a comprehensive framework to ensure that private sector organizations in Canada handle personal information responsibly and in accordance with individuals' privacy rights.

\section{Data Brokers practices and analysis}
Data Brokers practices and analysis are illustrated in Table \ref{tab:data}
\begin{longtblr}[
  caption = {Data Broker practices and analysis},
  label = {tab:data},
]{
  width = \linewidth,
  colspec = {Q[63]Q[69]Q[75]Q[260]Q[131]Q[338]},
  cells = {c},
  hlines,
  vlines,
}
\textbf{Data Broker Name} & \textbf{Year of Data Breach} & \textbf{Extent of Breach} & \textbf{Detailed Dataset Leaked}who are accountable for GDPR compliance (such as a Data P & \textbf{Remarks} & \textbf{Assumed GDPR Article Violation Reference Number \textbackslash{}\& Title}\\
\textbf{Social Data} & 2020 & Nearly 235 million profiles & Instagram, TikTok, and YouTube profiles with personal information and social media activity & exposed on a server without authentication or a password. & Art. 25 (Data Protection by Design and by Default), Art. 32 (Security of Processing)\\
\textbf{Inter- active Data} & 2020 & N/A & Full Social Security Numbers, dates of birth, all known physical addresses, email addresses, vehicle registrations, lines of credit, IP addresses & Data used by criminals for fraud and theft due to potential hack. & Art. 32 (Security of Processing), Art. 33 (Notification of a Personal Data Breach), Art. 34 (Communication of a Personal Data Breach)\\
\textbf{People Data Labs} & N/A & Over 1.2 billion records & thorough profiles that include email addresses, work histories, phone numbers, and social media profiles & Data exposure likely due to poor security by customers. & Art. 5 (Principles Relating to Processing of Personal Data), Art. 25 (Data Protection by Design and by Default), Art. 32 (Security of Processing)\\
\textbf{Lime Leads} & 2019 & Data on 49 million people & N/A (Specific dataset details not provided in the source) & Internal server not password-protected. & Art. 25 (Data Protection by Design and by Default), Art. 32 (Security of Processing)\\
\textbf{Exactis} & 2018 & Nearly 340 million people & Comprehensive personal information database & Exposed through an unsecure server. & Art. 25 (Data Protection by Design and by Default), Art. 32 (Security of Processing)\\
\textbf{Apollo} & 2018 & Billions of data points & Email addresses and possibly other related personal data & Data exposed due to hacking. & Art. 32 (Security of Processing), Art. 33 (Notification of a Personal Data Breach)\\
\textbf{Equif- ax} & 2017 & 147 million people & Names, addresses, Social Security Numbers, driver’s license numbers, credit card numbers, and other personal details & Information was stolen and used for brokering. & Art. 5 (Principles Relating to Processing of Personal Data), Art. 32 (Security of Processing), Art. 33 (Notification of a Personal Data Breach), Art. 34 (Communication of a Personal Data Breach)
\end{longtblr}

\begin{itemize}
    \item \textbf{Social Data:}
In 2020, a breach in Social Data exposed 235 million profiles from platforms such as Instagram and TikTok. Personal information and social media activity were stored unprotected on a server, potentially in violation of GDPR Articles 25 and 32 \cite{r22}.
\end{itemize}

\begin{itemize}
    \item \textbf{Interactive Data:}
in 2020, Interactive Data experienced a breach that compromised full social security numbers and personal information, resulting in criminal use. GDPR Articles 32, 33, and 34 could have been violated as a result of security flaws and a failure to properly notify affected individuals. 
\end{itemize}

\begin{itemize}
    \item \textbf{People Data Labs:}
people Data Labs experienced a data breach due to poor security, affecting over 1.2 billion records with detailed profiles. Potential GDPR violations include Articles 5, 25, and 32, which emphasize data processing principles, protection by design, and security measures\cite{r23}.
\end{itemize} 

\begin{itemize}
    \item \textbf{LimeLeads:}
LimeLeads' 2019 data breach exposed information on 49 million people from an unprotected internal server. GDPR Articles 25 and 32 could have been violated, exposing gaps in data protection measures\cite{r24}.
\end{itemize}  

\begin{itemize}
    \item \textbf{Exactis:}
A breach at Exactis in 2018 resulted in the complete personal data of 340 million people being exposed via an insecure server. Articles 25 and 32 of the GDPR, which highlight the importance of data protection and secure processing, may be violated.
\end{itemize}

\begin{itemize}
    \item \textbf{Apollo:}
In 2018, Apollo experienced a data breach involving billions of data points, including email addresses and possibly other personal information. The exposure was caused by hacking activities. This incident may have violated GDPR Articles 32 and 33, which emphasize lapses in personal data processing security and the requirement for timely notification of a personal data breach.
as processing security.
\end{itemize}

\begin{itemize}
    \item \textbf{Equifax:}
The Equifax data breach in 2017 exposed 147 million people's personal information, including names, addresses, and Social Security numbers. This incident most likely violated GDPR Article 5 on data processing principles and Article 32 on processing security. Articles 33 and 34, which require prompt notification of a breach, would also be relevant. This incident highlights the critical importance of strong data protection measures to prevent unauthorized access and protect individuals' privacy.
\end{itemize}

\section{Proposed framework}
\begin{figure}[htbp]
\centering
  \includegraphics[width=0.7\columnwidth]{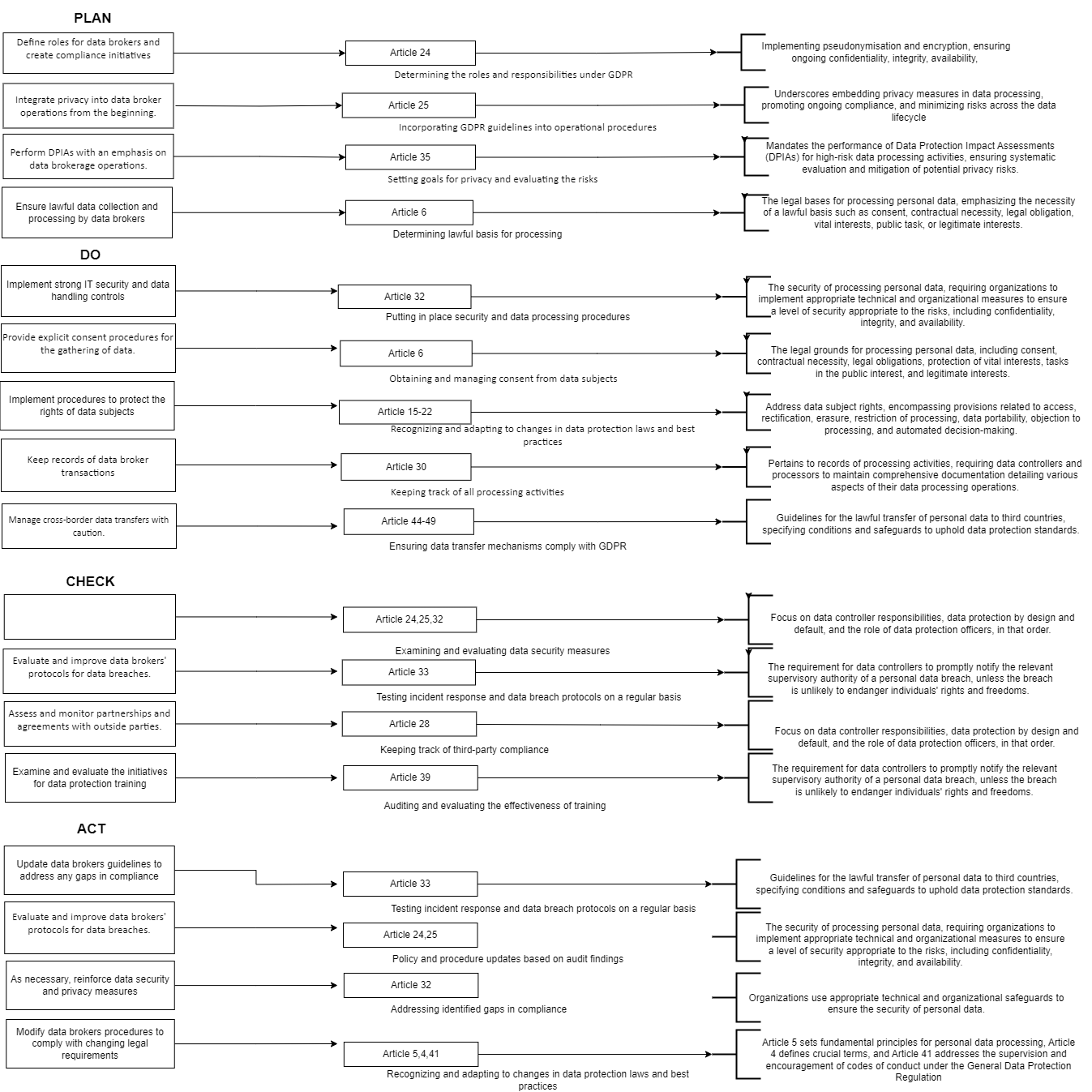}
  \caption{Proposed framework/strategy}~\label{fig:figure4}
\end{figure}

The conceptual framework we are planning to propose is illustrated in \ref{fig:figure4}. 

     \textbf{Plan Phase:}
    \begin{itemize}
      \item \textbf{Define roles for Data Brokers and create compliance initiatives:} 
      Data Brokers should identify individuals within their organization who are accountable for GDPR compliance (such as a Data Protection Officer) and form a data protection team or department. They must develop comprehensive initiatives that comply with GDPR requirements, such as privacy training programs and internal policies (Article 24). 
      
      \textbf{Example:} "DataWorld Inc." appoints a Data Protection Officer (DPO) and forms a GDPR compliance team to develop a privacy program that includes mapping out data flows, assessing risks, and developing a GDPR compliance checklist.
      
      \textbf{Reference:} Article 24 requires organizations to implement data protection policies. DataWorld Inc. would document the delegation of responsibilities, reporting structures, and staff training requirements.

      \item \textbf{Integrate privacy into Data Broker operations from the beginning:} 
      Incorporating 'Data Protection by Design and by Default' (Article 25) requires Data Brokers to consider privacy during the design stage of any new product, service, or data processing activity, and to ensure that personal data is processed with the highest privacy settings by default.
 
      \textbf{Example:} "Consumer Insights Ltd." launches a new project to collect consumer data for market research. They include privacy controls like data anonymization and pseudonymization from the start. 
      
      \textbf{Reference:} Article 25 requires that data protection be built into the processing, not as an afterthought, and Consumer Insights Ltd. incorporates these safeguards into their initial system specifications.

      \item \textbf{Perform DPIAs with an emphasis on Data Brokerage operations:} 
      When beginning new data processing activities that may jeopardize individuals' rights and freedoms, Data Brokers should conduct Data Protection Impact Assessments (Article 35). This procedure aids in the early detection and mitigation of risks. 
      
      \textbf{Example:} Before launching a new data analytics platform, "DataWorld Inc." conducts a DPIA to assess how the platform could affect individual privacy and identifies measures to mitigate those risks. 
      
      \textbf{Reference:} Before launching a new data analytics platform, "DataWorld Inc." conducts a DPIA to assess how the platform may affect individual privacy and identifies mitigation measures.

      \item \textbf{Ensure lawful data collection and processing by Data Brokers:} 
      Data Brokers must have a legal basis for collecting and processing personal data, whether through consent, contract necessity, legal obligation, or other means specified in Article 6.
 
      \textbf{Example:} Consumer Insights Ltd." ensures that all data collection campaigns have clear, understandable consent forms that meet GDPR lawfulness standards, including opt-in check boxes for various types of data processing.

      \textbf{Reference:} Article 6 specifies the legal basis for processing personal data. This requirement would be met by Consumer Insights Ltd.'s consent mechanism.
\end{itemize}

   \textbf{Do Phase:}
    \begin{itemize}
      \item \textbf{Deploy robust IT security and data handling controls:} 
      Encryption, access control, and pseudonymization are examples of technical and organizational measures to secure personal data (Article 32). 
      
      \textbf{Example:}  "DataWorld Inc." uses end-to-end encryption for data in transit and at rest, performs regular penetration tests, and keeps a secure data processing environment.

      \textbf{Reference:} Article 32 requires the implementation of appropriate security measures, and DataWorld Inc.'s actions would be directly in compliance.
\end{itemize}

       \begin{itemize}
      \item \textbf{Establish clear consent mechanisms for data collection:} 
      Article 7 requires Data Brokers to ensure that consent is freely given, specific, informed, and unambiguous. They should create transparent mechanisms that make it simple for people to give and withdraw consent.
      
      \textbf{Example:}  "Consumer Insights Ltd." uses a digital consent management platform that allows users to provide, review, and withdraw their consent for different processing activities.
      
      \textbf{Reference:} Article 7 emphasizes the importance of a clear and affirmative consent mechanism, which would be provided by the consent management platform.
\end{itemize}

       \begin{itemize}
      \item \textbf{Implement processes to address data subjects' rights:}
      Data Brokers must make it easier for data subjects to exercise their rights, from access to deletion (Articles 15 to 22). This includes developing request processes that are both efficient and user-friendly. 
      
      \textbf{Example:}  DataWorld Inc." creates an online portal where individuals can easily access their data, request corrections, or exercise their right to be forgotten.

      \textbf{Reference:} Articles 15 to 22 cover data subjects' rights, and the portal would make it easier to exercise these rights.
\end{itemize}

\begin{itemize}
      \item \textbf{Maintain comprehensive records of Data Broker transactions:} 
      Keeping detailed records of processing activities (Article 30) is critical for demonstrating GDPR compliance.
      
      \textbf{Example:}  "Consumer Insights Ltd." keeps detailed records of all data processing activities, including data sources, consent records, and processing purposes, using data processing and inventory management software

      \textbf{Reference:}Article 30 requires the keeping of records of processing activities, which the software assists with.
\end{itemize}

\begin{itemize}
      \item \textbf{Manage cross-border data transfers with due diligence:}
      Data Brokers must ensure that appropriate safeguards are in place when transferring data outside the EU (Articles 44 to 49). 
      
      \textbf{Example:}  DataWorld Inc." ensures that when transferring data outside the EU, it only partners with entities in countries deemed to provide adequate protection or uses Standard Contractual Clauses. Data Brokers must ensure that appropriate safeguards are in place when transferring data outside the EU (Articles 44 to 49).
      
      \textbf{Reference:}Articles 44 to 49 define the conditions for transferring personal data outside the EU, and DataWorld Inc.'s practices must comply with these requirements. 
\end{itemize}
 
 \textbf{Check Phase:} 
 Data Brokers should review and evaluate their data protection measures on a regular basis to ensure they are effective and in accordance with Articles 24, 25, and 32.
 
\begin{itemize}
    \item "DataWorld Inc." and "Consumer Insights Ltd." conduct regular reviews and evaluations to assess the effectiveness of their GDPR compliance measures. They may test their incident response plans with mock data breaches and use third-party auditors to review third-party vendor compliance.

\end{itemize}

Evaluate and improve Data Brokers' data breach protocols: In accordance with Article 33, incident response plans and data breach notifications should be tested on a regular basis. 

Examine and monitor external partnerships and agreements: Monitoring third-party GDPR compliance, as outlined in Article 28, is critical to ensuring that partners and vendors also adhere to data protection standards.

Examine and evaluate the following data protection training initiatives: An audit of the effectiveness of data protection training helps ensure that employees understand the GDPR (Article 39).

\textbf{Act Phase:}
 
\begin{itemize}
    \item Based on their reviews, both companies would take corrective actions. If gaps are discovered, they may need to update their training programs, revise their data handling procedures, or strengthen their IT security measures. They would also keep up with changes in legal requirements, adjusting their policies and practices as needed.
\end{itemize}

To address any gaps in compliance, update Data Brokerage guidelines: According to Articles 24 and 25, Data Brokers should update their policies and procedures based on the findings of audits and reviews in order to close any compliance gaps. 

Strengthen data security and privacy safeguards: Article 32 requires that any identified weaknesses or compliance gaps in security measures be addressed as soon as possible. 

Adapt Data Brokerage procedures to changing legal requirements: Data Brokers must be flexible in order to adapt to changes in data protection laws and best practices. They must modify their procedures as needed to comply with the GDPR's general principles and accountability requirements (Recital 4, Articles 5, 40, 41).
By adhering to the PDCA cycle, Data Brokers can ensure continuous improvement in their data protection practices, resulting in improved GDPR compliance and better personal data protection.

%\chapter{\textbf{PART III}}

\section{Conclusion}
In conclusion, the practices of Data Brokers present multifaceted implications spanning privacy, ethics, security, and regulatory compliance. The collection of extensive personal information without explicit consent raises significant privacy concerns, and the influential role of Data Brokers in decision-making prompts ethical considerations. Data accuracy is pivotal to avoiding financial and social consequences for individuals, while the security risks associated with data breaches underscore the necessity for robust cybersecurity measures within the industry. It's imperative that Data Brokers organization should adopt data privacy and governance frameworks either developed in-house or adopted from the industry.
Regulatory compliance, particularly with frameworks like GDPR will also help ensure transparency and protect individual rights.
In light of these implications, recommendations include reinforcing ethical standards in data collection, prioritizing accuracy to mitigate financial and social impacts on individuals, and implementing stringent cybersecurity measures to safeguard against data breaches. Collaboration with regulatory bodies to enhance and enforce privacy laws is essential, emphasizing the need for a global approach to data protection. Additionally, fostering increased data literacy among individuals can empower them to understand and assert control over the use of their personal information. Ultimately, balancing the benefits of data-driven insights with ethical considerations and individual rights is pivotal for the responsible and sustainable evolution of the Data Broker industry.

%Bibliography

\clearpage
\newpage % Start a new page for the appendix.
\bibliographystyle{unsrt}
\bibliography{bibs_file.bib}

\end{document}